\documentclass[10pt,  twocolumn, twoside, letterpaper, conference, final]{IEEEtran}
\IEEEoverridecommandlockouts
\usepackage{amsmath}

\usepackage{examplep}  
\usepackage{graphicx}
\usepackage{cite}
\usepackage{algorithmicx}
\usepackage[ruled]{algorithm}
\usepackage{algpseudocode}
\usepackage{mathtools}
\usepackage{dsfont}
\usepackage{amsfonts}
\usepackage{amssymb}
\usepackage{balance}
\usepackage{booktabs,tabularx}
\usepackage{multirow}
\usepackage{textcomp}
\usepackage[usestackEOL]{stackengine}
\usepackage[hyphens]{url}
\usepackage{pict2e}
\usepackage{color}
\usepackage{enumerate}
\usepackage{float}
\usepackage{mathrsfs}
\usepackage{mathtools}
\usepackage{subcaption}
\usepackage{tikz}
\usetikzlibrary{shapes.arrows,decorations.pathreplacing,decorations.markings,shapes.geometric}
\tikzset{radiation/.style={{decorate,decoration={expanding waves,angle=0,segment length=2pt}}}} 
\usetikzlibrary{positioning}
\usepackage{graphicx}
\usepackage[normalem]{ulem}

\DeclareMathAlphabet\mathbfcal{OMS}{cmsy}{b}{n}

\makeatletter
\def\blfootnote{\gdef\@thefnmark{}\@footnotetext}
\makeatother

\begin{document}
\title{Comparison of Short Blocklength Slepian-Wolf Coding for Key Reconciliation}
\IEEEoverridecommandlockouts
%
\author{\IEEEauthorblockN{Mahdi Shakiba-Herfeh,
and Arsenia Chorti}
\IEEEauthorblockA{ETIS UMR8051, CY Universite, ENSEA, CNRS, F-95000, Cergy, France\\
Email: \{mahdi.shakiba-herfeh, arsenia.chorti\}@ensea.fr}
}
\maketitle
\begin{abstract} 


We focus  Slepian-Wolf (SW) coding in the short blocklength for reconciliation in secret key generation and physical unclonable functions. In the problem formulation, two legitimate parties wish to generate a common secret key from a noisy observation of a common random source in the presence of a passive eavesdropper. We consider three different families of codes for key reconciliation. The selected codes show promising performances in information transmission in the short blocklength regime. We implement and compare the performance of different codes for SW reconciliation in the terms of reliability and decoding complexity.

\end{abstract}

\section{Introduction} \label{sec:intro}

Wireless communication technologies have become an essential part of our everyday lives.  As  more  and  more  data  are  being  transmitted  over  wireless  channels,  besides  the reliability of the transmitted information, guaranteeing the security of information transferred has become a challenging issue. Physical layer security (PLS) has been investigated in the recent years to provide the security in information exchanged information by exploiting the properties of the physical layer \cite{Chorti16, shakibaPLS,Ersi_Context_aware}.

In the context of PLS, in this paper we consider the problem of coding for secret key generation (SKG) from a common source of randomness observed by two communication parties\footnote{The proposed codes can be further applied in the reconciliation step in the context of physical unclonable functions \cite{MiroEurasip}.}. In the problem formulation, the observation of two legitimate parties, who wish agree on a common secret key, differ due to noise or other factors. The central idea of reconciliation coding for SKG is that the parties should be able to correct the mismatches, by sending some helper data trough the public communication channel. Of course, the helper data can also help an eavesdropper (Eve) to have a better guess about the respective measurements. 

Therefore, in coding for SKG, the target is to design a coding scheme which requires the minimum amount of helper data (information leakage), while guaranteeing that the legitimate parties will be able to reconcile the mismatches with a high probability (reliability). The coding for key reconciliation is also referred to as Slepian-Wolf (SW) coding in the literature. Motivated by the increasing importance of low latency communications in 
recent years, we focus on the short blocklength SW coding in this contribution.

The problem of coding for SW reconciliation has been studied in the literature for long blocklengths. The SKG based on polar codes have been studied in \cite{coding1,coding7}. In \cite{coding3,coding2}, Bose Chaudhuri Hocquenghem (BCH) codes are utilized for SW coding. Also LDPC codes have been proposed and their performance has been analysed in \cite{coding5}. However, as most of the work on coding for SW settings only consider long blocklengths, respective results for the short blocklength have not been reported.

In this paper we investigate coding for reconciliation where the two parties have generated a binary random sequence from their observations of a common random source. As an example, the possible source of random variables can be the wireless channel state or the challenge-response of physical unclonable functions (PUF). We implement and compare the reliability / complexity performance of polar, LDPC and BCH codes in the short blocklength regime. All the selected codes show promising performance in information transmission in the short blocklength. 

The paper is organized as follows. In Section \ref{sec:Sys_model}, we describe the system model. In Section \ref{sec:codes}, we explain the details of the codes' implementations, and in Section \ref{sec:results} we present the numerical results and compare the performance of the codes investigated. Finally, we conclude the paper in Section \ref{sec:conclusion}.

\section{System Model}\label{sec:Sys_model}

In our system model, we assume Alice and Bob observe binary sequences $Y_A, Y_B$ of length $n$ by quantizing their local measurements of a common source of randomness, represented by $X_A$ and $X_B$, respectively. At present, for the sake of simplicity, we assume that a passive adversary, referred to as Eve, is unable to obtain any information about the observed sequences.

We assume that the generated binary sequences are independent and identically distributed (i.i.d.) with bits at any index having equal probabilities of being $0$ or $1$, i.e.,
\begin{align}
&Pr\left(Y_A[i]=0\right)=Pr\left(Y_B[i]=0\right)=0.5 \\
&Pr\left(Y_A[i]\neq Y_B[i]\right) = p, i=1,\ldots,n.
\end{align}
Alice and Bob attempt to agree on a secret key $K$ of length $k$ that must be drawn uniformly from a length $n$ binary tuple $\mathcal{K} = \{0, 1\}^k$. To this end, Alice sends her syndrome $S$, of length $n-k$, through a public channel in order to help Bob to obtain an estimate $\hat{Y}_A$ of her sequence $Y_A$. Additionally, Bob and Eve can listen to the public channel without error.

\begin{figure}[t]
\centering
\includegraphics[ width=0.5\textwidth]{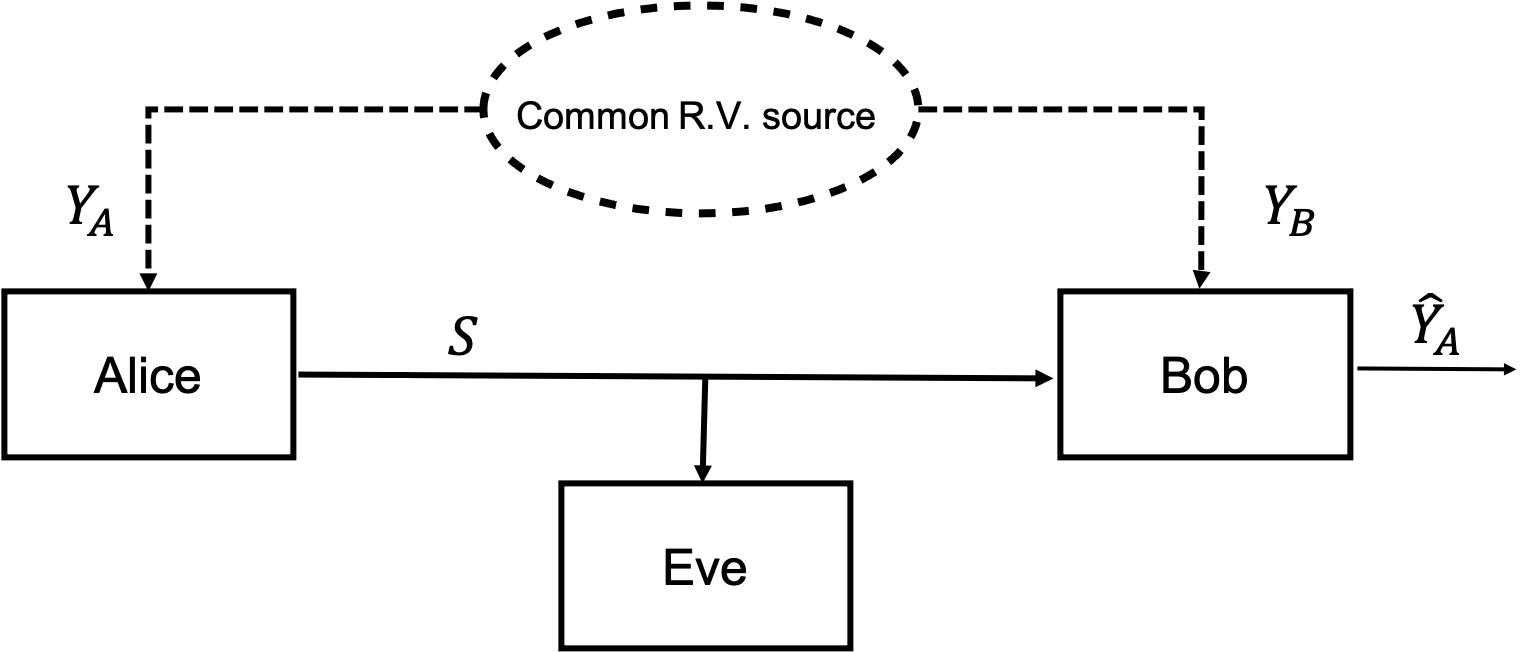}
\caption{The system model.}
\label{fig:SysModel}
\end{figure} 

The code rate is defined as $R = \frac{k}{n}$ and the frame error rate (FER) is defined as the probability that Bob's estimation of $Y_A$ is erroneous $Pr(Y_A \neq \hat{Y}_A)$. In this set-up, Bob first estimates the sequence $\hat{Y}_A$ and then, Alice and Bob extract the common key $K$ independently using privacy amplification. (Fig. \ref{fig:SysModel}).
Motivated by the promising performance of LDPC \cite{LDPCList}, polar \cite{PolarList}, and BCH codes \cite{BCHList} with list decoding in short blocklengths for information transformation applications, in this paper we implement and compare against the upper bound on SKG rates in \cite{SKG_upperbound}, when these three families of SW decoders are employed.

\section{Codes Implementation}\label{sec:codes}

\subsection{LDPC codes with ordered statistic decoding}
LDPC codes are powerful error correcting codes that can approach the Shannon limit at very large blocklengths. They exhibit promising performance for a variety of communication channels \cite{Mahdi_LDPCdesign2, Mahdithesis,MahdiCIC}.  However, in general, they do not work well in short blocklengths. To address such shortcomings, LDPC codes enhanced with ordered statistic decoding (OSD) is one of the techniques proposed to achieve near maximum likelihood (ML) efficiency for short blocklengths \cite{LDPCList}.
The main concept behind OSD is that we start by selecting the $\tilde k$ most reliable independent positions, where $\tilde k$ is the code rank. Afterwards, we make hard decisions on the value of the selected bits, based on the log-likelihood ratios (LLR). Then, by flipping the values of up to $t$ bits among them, we produce a candidate list of codewords. Finally, we choose the most likely codeword by performing an ML search in the list \cite{OSD2}. The size of the candidate OSD list increases with respect to $t$ as $\sum^t_{i=0} \binom ki$.

\begin{figure}[h]
\centering
\includegraphics[ width=0.4\textwidth]{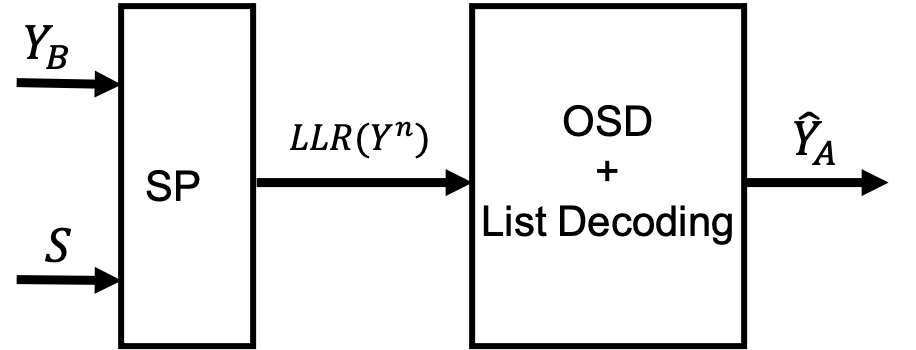}
\caption{Decoder structure for LDPC codes with ordered statistics decoding.}
\label{fig:LDPC}
\end{figure}

In our implementation, Alice sends her syndrome $S = Y_A H^t$, where $H^t$ is the transpose of the parity check matrix. At the receiver side, Bob first feeds $Y_B$ and $S$ to the sum-product (SP) algorithm to generate soft information of the bits (i.e., the LLRs). Then the LLRs are passed to the OSD block to estimate $\hat{Y}_A$ (Fig. \ref{fig:LDPC}). In the iterative version of LDPC code with OSD, to enlarge the candidate list of codewords, the LLRs after each iteration of SP algorithm are passed to OSD, and at the end the OSD block pick the most likely codeword. The size of the candidate list of codewords in iterative scheme is at most $itr$ times larger than the non-iterative scheme, where the $itr$ denotes the number of iterations of the SP algorithm. 

\subsection{Polar codes with list decoding}
Polar codes are linear block error correcting codes that can provably achieve the capacity of a binary-input discrete memoryless channel as the code length goes to infinity. Using successive cancellation list decoding, which holds a list of most likely decoding paths, can significantly improve the performance of polar codes in finite blocklengths. The cyclic redundancy check (CRC) will boost the performance of list decoding even further. The CRC aids the decoder in selecting the correct decoding route from a list of options, even if it is not the most likely one. \cite{PolarList}.

\begin{figure}[t]
\centering
\includegraphics[ width=0.25\textwidth]{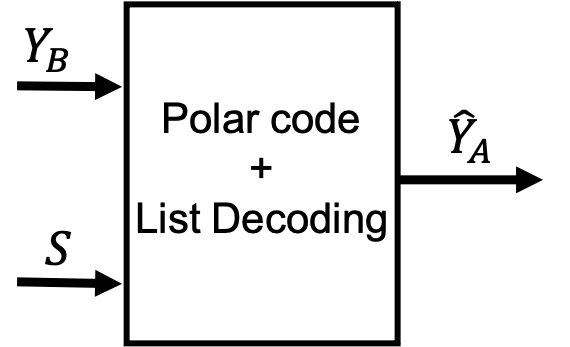}
\caption{Decoder structure for polar codes with list decoding.}
\label{fig:polar}
\end{figure}

In our implementation (similar to \cite{Polar_source}), Alice encodes her sequence $Y_A$ as $U = Y_A G_n$, where $G_n = \big(\begin{smallmatrix}
  1 & 0\\
  1 & 1
\end{smallmatrix}\big)^{\otimes n}$ is the encoder matrix as defined in \cite{Polar_source}. Alice sends the syndrome $S$ which contains $S_1$ and CRC bits with length $l$. $S_1$ has length $n-k-l$ and contains high-entropy bits of $U$ as follows
\begin{equation}\label{eq:Polar_encoder}
H(U[i]|Y_A,U^{i-1})\geq H(U[j]|Y_A,U^{j-1}),\; 1\leq i,j\leq n,
\end{equation}
where $i$ is the position of transmitted bits and $H(\cdot)$ denotes entropy. Therefore, the actual rate of the  polar code is $R=\frac{k+l}{n}$. On the other side, Bob applies CRC-aided successive cancellation list decoding to estimate $\hat Y_A$. In the list decoding scheme, Bob tracks $L$ decoding paths simultaneously, where $L$ is the list size. Finally, the decoder picks the most likely codeword which also satisfies the CRC condition among the $L$ paths (Fig. \ref{fig:polar}). Note that the complexity of list decoding polar coding grows linearly with the list size. 

\subsection{BCH codes with list decoding}
BCH codes are a class of cyclic error-correcting codes constructed by polynomials over a finite field. One of the main features of BCH codes during code design is the number of guaranteed correctable error bits. A binary BCH code is defined by $(n_{BCH} ,k_{BCH} ,t_{BCH} )$, where $n_{BCH}=2^w-1$ is the blocklength, $k_{BCH}$ is the message length, and $t_{BCH}$ represents the number of guaranteed correctable error bits. To improve their error correcting capability, BCH codes can be armed with list decoding.

\begin{figure}[t]
\centering
\includegraphics[ width=0.5\textwidth]{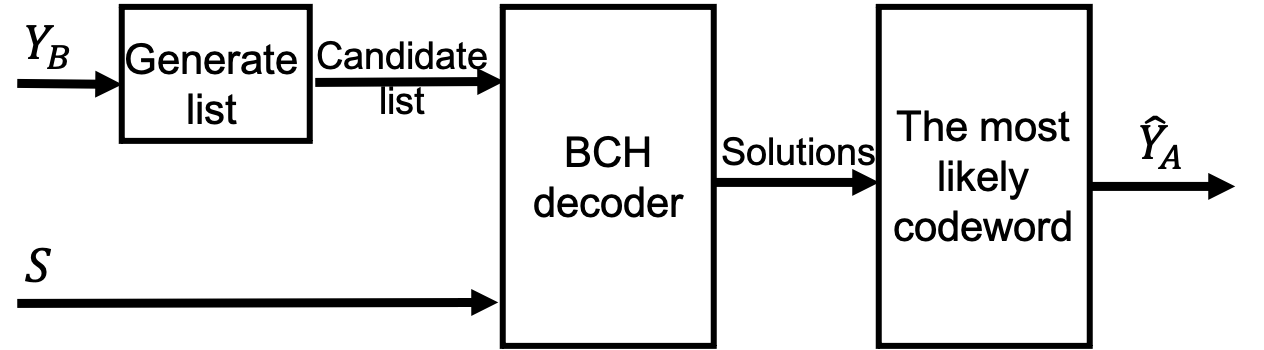}
\caption{Decoder structure for BCH codes with list decoding.}
\label{fig:BCH}
\end{figure}

In our implementation of list decoding, Alice calculates the syndromes as $S = Y_A H^t$, where $H$ is the parity check matrix of the BCH code, and transmits it through the public channel. On the other side, Bob, first generates a candidate list by flipping up to $t$ bits of the measured sequence $Y_B$. After feeding the candidate list to the BCH decoder, the BCH decoder finds a possible solution for each element of the candidate list and picks the solution which is the most likely codeword with respect to the measured sequence $Y_B$ (Fig. \ref{fig:BCH}). In this implementation of list decoding, the size of the list increases with respect to $t$ as $\sum^t_{i=0} \binom ni$. To reduce the size of list, without a significant degrade in the performance of decoder some heuristic methods exist in the literature \cite{BCH_Cho}.
 
\begin{figure*} [!h]
  \includegraphics[width=\textwidth,height=8.5cm]{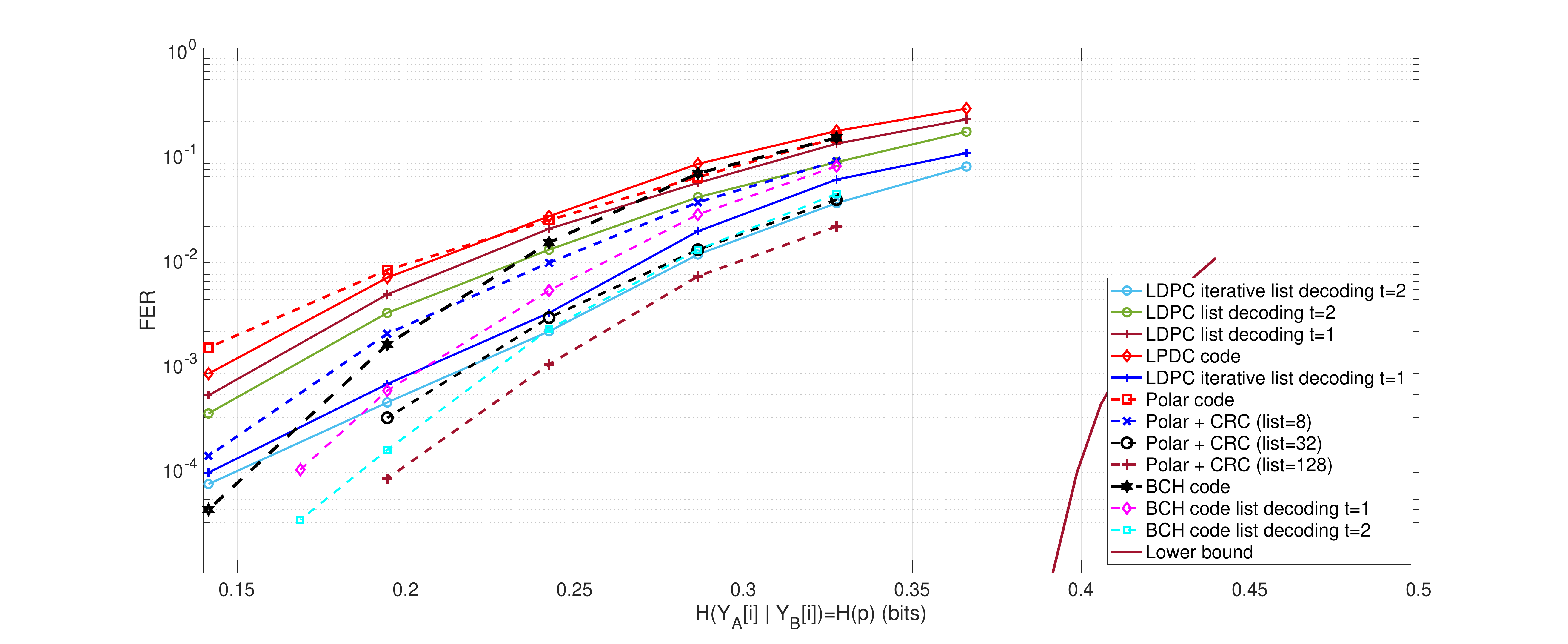}
  \caption{Comparison of FER performance of codes with the lower bound in \cite{SKG_upperbound} for $n=128$.}\label{Fig:FERN128}
\end{figure*}

\begin{figure*} [!h]
  \includegraphics[width=\textwidth,height=8.5cm]{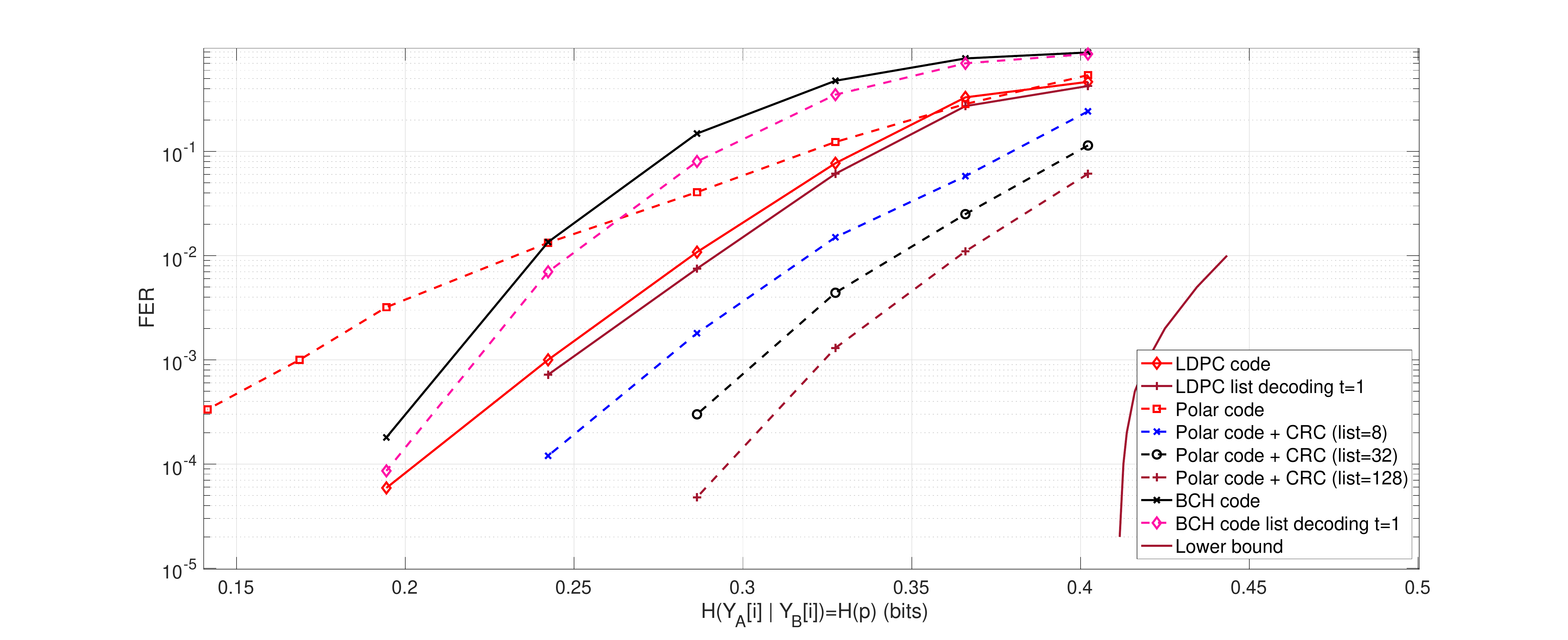}
  \caption{Comparison of FER performance of codes with the lower bound in \cite{SKG_upperbound} for $n=512$.}\label{Fig:FERN512}
\end{figure*}

\section{Numerical results}\label{sec:results}
In this section, we provide the FER performance of the aforementioned codes in the SKG setting for 128 and 512 bits. We also compare the FER performance of codes with the information theoretical finite length upper bound reported in \cite{SKG_upperbound}, which translates to a lower bound in the FER.

For the first instance, we consider blocklength $128$ bits. We implement a $(n,k)=(128,75)$ polar code with 11 bits CRC and a $(127,64,10)$ BCH code. Also, we pick regular $(3,6)$ LDPC codes with length 128 bits\footnote{We note that, the variable and check nodes degree distributions can be optimized to improve the performance of LDPC codes \cite{Mahdi_LDPCdesign}, however the degree optimization of the LDPC codes is out of the scope of this work.}. The maximum number of iterations for SP algorithm is set to 50. In Fig. \ref{Fig:FERN128}, the FER performances of half rate codes with $128$ bits blocklength are depicted (i.e., the key length after privacy amplification is $64$) and compared to the upper bound reported in \cite{SKG_upperbound}. Note that the upper bound becomes inaccurate for short blocklength while their accuracy improves as the blocklength increases. That is the reason we observe comparatively a large gap between the upper bound and the codes' FER performances for $n=128$ \cite{coding7}.

As it is depicted in Fig. \ref{Fig:FERN128}, the polar code with list size $128$, second order BCH list decoding code, and LDPC iterative list decoding for $t=2$ outperform and show almost similar performance at FER $~10^{-3}$. Moreover, The results show that, while traditional polar codes perform poorly in short blocklengths, their performance can be significantly improved by equipping them with list decoding. For example, the polar code with list size $128$, provides around two orders lower FER compared to the classical polar code at $H(Y_A[i]|Y_B[i])=0.1944$. However, this improvement comes at the cost of $128$ times more decoding complexity. We also observe that LDPC iterative list decoding outperforms the non-iterative one. Furthermore, list decoding also improves the performance of the BCH and LDPC codes, but the improvement with respect to the additional complexity is not as notable as in the case of the polar code.

Furthermore, to generate keys that can be used with standard block ciphers, e.g., AES-128 or AES-256, the SKG process is assumed to run with blocklength 512 and rate half.  As it is demonstrated in Fig. \ref{Fig:FERN512}, we observe a significant improvement in the performance of polar codes by using list decoding. On the other hand, we do not observe a noticeable improvement by using list decoding size $t = 1$ for LDPC and BCH code for this blocklength. We conclude this observations as the strength of list decoding with a fixed size as the blocklength increases, diminishes. We do not generate higher list decoding order for LDPC and BCH codes due to their high decoding complexity. Also, at blocklength $n=512$, the gap between the FER of the polar code with list size 128 and the lower bound is less than the case $n=128$. We posit that one of the reasons is that the lower bound at this length is tighter than the first instance.

In the simulations for short blocklength, we observe that although the considered implementation of list decoding for LDPC and BCH codes improves the performance of the code in short blocklength ($\sim$100 bits). However, the improvement diminishes as we increase the blocklength, and to obtain enhancement in larger blocklength, we need to increase the order of the list decoding, which adds the complexity of decoders significantly. For these schemes utilizing heuristic methods, to modify the list size, can effectively reduce the complexity. On the other hand, we observe that polar codes with list decoding provide a significant improvement for medium size blocklength as well as short blocklength.

\section{Conclusion}\label{sec:conclusion}
In this paper, we consider coding for Slepian-Wolf  key reconciliation in the short blocklength regime. We assume that two legitimate parties observe binary sequences, generated from measurements of a common random source. We implement three different families of codes, including LDPC codes, BCH codes, and Polar codes to detect and correct the mismatches in the noisy observations. Our simulations for  blocklengths of $n=128$ and $n=512$ bits, considering both reliability and decoding complexity, demonstrate that polar codes with a list size of 128 outperforms all other codes.
\section*{Acknowledgements}
This work has been supported by the ANR PRCI project ELIOT (ANR-18-CE40-0030 / FAPESP 2018/12579-7) and the INEX Paris Universite projects PHEBE and eNiGMA.
\balance
\bibliographystyle{IEEEtran}

\bibliography{refs}

\begin{thebibliography}{10}
\providecommand{\url}[1]{#1}
\csname url@samestyle\endcsname
\providecommand{\newblock}{\relax}
\providecommand{\bibinfo}[2]{#2}
\providecommand{\BIBentrySTDinterwordspacing}{\spaceskip=0pt\relax}
\providecommand{\BIBentryALTinterwordstretchfactor}{4}
\providecommand{\BIBentryALTinterwordspacing}{\spaceskip=\fontdimen2\font plus
\BIBentryALTinterwordstretchfactor\fontdimen3\font minus
  \fontdimen4\font\relax}
\providecommand{\BIBforeignlanguage}[2]{{%
\expandafter\ifx\csname l@#1\endcsname\relax
\typeout{** WARNING: IEEEtran.bst: No hyphenation pattern has been}%
\typeout{** loaded for the language `#1'. Using the pattern for}%
\typeout{** the default language instead.}%
\else
\language=\csname l@#1\endcsname
\fi
#2}}
\providecommand{\BIBdecl}{\relax}
\BIBdecl

\bibitem{Chorti16}
A.~Chorti, C.~Hollanti, J.-C. Belfiore, and H.~V. Poor, ``Physical layer
  security: A paradigm shift in data confidentiality,'' in \emph{Physical and
  Data-Link Security Techniques for Future Communication Systems}.\hskip 1em
  plus 0.5em minus 0.4em\relax Cham: Springer International Publishing, 2016,
  pp. 1--15.

\bibitem{shakibaPLS}
M.~Shakiba-Herfeh, A.~Chorti, and H.~V. Poor, ``Physical layer security:
  authentication, integrity and confidentiality,'' \emph{arXiv:2001.07153},
  2020.

\bibitem{Ersi_Context_aware}
A.~Chorti, A.~N. Barreto, S.~Kopsell, M.~Zoli, M.~Chafii, P.~Sehier,
  G.~Fettweis, and H.~V. Poor, ``Context-aware security for {6G} wireless the
  role of physical layer security,'' \emph{arXiv:2101.01536}, 2021.

\bibitem{MiroEurasip}
M.~Mitev, A.~Chorti, M.~Reed, and L.~Musavian, ``Authenticated secret key
  generation in delay-constrained wireless systems,'' \emph{EURASIP J Wireless
  Com Network}, vol. 122, 2020.

\bibitem{coding1}
B.~{Chen \textit{et al.}}, ``A robust {SRAM-PUF} key generation scheme based on
  polar codes,'' in \emph{Proc. IEEE Global Commun. Conf.}, 2017, pp. 1--6.

\bibitem{coding7}
H.~{Hentilä \textit{et al.}}, ``On polar coding for finite blocklength secret
  key generation over wireless channels,'' in \emph{Proc. IEEE Int. Conf.
  Acoustics, Speech and Signal Process. (ICASSP)}, 2020, pp. 5265--5269.

\bibitem{coding3}
R.~Maes, A.~Van~Herrewege, and I.~Verbauwhede, ``{PUFKY}: A fully functional
  {PUF}-based cryptographic key generator,'' in \emph{Cryptographic Hardware
  and Embedded Systems -- CHES}, 2012, pp. 302--319.

\bibitem{coding2}
S.~{Forchhammer}, M.~{Salmistraro}, K.~J. {Larsen}, X.~{Huang}, and H.~V.
  {Luong}, ``Rate-adaptive {BCH} coding for slepian-wolf coding of highly
  correlated sources,'' in \emph{2012 Data Compression Conference}, 2012, pp.
  237--246.

\bibitem{coding5}
D.~{Elkouss}, J.~{Martinez}, D.~{Lancho}, and V.~{Martin}, ``Rate compatible
  protocol for information reconciliation: An application to {QKD},'' in
  \emph{Proc. IEEE Inf. Theory Workshop (ITW 2010, Cairo)}, 2010, pp. 1--5.

\bibitem{LDPCList}
W.~{Zhou} and M.~{Lentmaier}, ``Improving short-length {LDPC} codes with a
  {CRC} and iterative ordered statistic decoding : (invited paper),'' in
  \emph{53rd Annual Conf. Inf. Sciences Systems (CISS)}, 2019, pp. 1--6.

\bibitem{PolarList}
I.~{Tal} and A.~{Vardy}, ``List decoding of polar codes,'' \emph{IEEE Trans.
  Inf. Theory}, vol.~61, no.~5, pp. 2213--2226, 2015.

\bibitem{BCHList}
M.~P.~C. {Fossorier} and {Shu Lin}, ``Soft-decision decoding of linear block
  codes based on ordered statistics,'' \emph{IEEE Trans. Inf. Theory}, vol.~41,
  no.~5, pp. 1379--1396, 1995.

\bibitem{SKG_upperbound}
H.~{Tyagi} and S.~{Watanabe}, ``Converses for secret key agreement and secure
  computing,'' \emph{IEEE Trans. Inf. Theory}, vol.~61, no.~9, 2015.

\bibitem{Mahdi_LDPCdesign2}
{M}.{Shakiba-Herfeh}, A.~K. {Tanc}, and T.~M. {Duman}, ``{LDPC} code design for
  fast fading interference channels,'' in \emph{2018 IEEE International
  Conference on Communications (ICC)}, 2018, pp. 1--6.

\bibitem{Mahdithesis}
M.~S. Herfeh, ``Code design for interference channels,'' Ph.D. dissertation,
  bilkent university, 2019.

\bibitem{MahdiCIC}
M.~{Shakiba-Herfeh}, A.~K. {Tanc}, and T.~M. {Duman}, ``Ldpc codes for
  interference channels in the primary decodes cognitive regime,'' \emph{IEEE
  Wireless Communications Letters}, vol.~8, no.~4, pp. 1187--1190, 2019.

\bibitem{OSD2}
Y.~{Wu} and C.~N. {Hadjicostis}, ``Soft-decision decoding using ordered
  recodings on the most reliable basis,'' \emph{IEEE Trans. Inf. Theory},
  vol.~53, no.~2, pp. 829--836, 2007.

\bibitem{Polar_source}
E.~{Arika{n}}, ``Source polarization,'' in \emph{IEEE Int. Symp. Inf. Theory},
  2010.

\bibitem{BCH_Cho}
J.~{Cho} and W.~{Sung}, ``Efficient software-based encoding and decoding of bch
  codes,'' \emph{IEEE Trans. Computers}, vol.~58, no.~7, pp. 878--889, July
  2009.

\bibitem{Mahdi_LDPCdesign}
M.~{Shakiba-Herfeh}, A.~K. {Tanc}, and T.~M. {Duman}, ``{LDPC} code design for
  fading interference channels,'' \emph{IEEE Trans. Veh. Technol.}, vol.~68,
  no.~3, pp. 2374--2385, 2019.

\end{thebibliography}

\end{document}